\documentclass[conference]{IEEEtran}
\IEEEoverridecommandlockouts
\usepackage{cite}
\usepackage{amsmath,amssymb,amsfonts}
\usepackage{algorithm}
\usepackage{algpseudocode}
\usepackage{graphicx}
\usepackage{subcaption}
\usepackage{textcomp}
\usepackage{xcolor}
\def\BibTeX{{\rm B\kern-.05em{\sc i\kern-.025em b}\kern-.08em
    T\kern-.1667em\lower.7ex\hbox{E}\kern-.125emX}}
\begin{document}

\title{\huge A Novel Deep Reinforcement Learning Method for Computation Offloading in Multi-User Mobile Edge Computing with Decentralization
}

\author{
    \IEEEauthorblockN{Nguyen Chi Long$^\ast$, Trinh Van Chien$^\ast$,   Ta Hai Tung$^\ast$, Van Son Nguyen$^\xi$, \\ Trong-Minh Hoang$^\eta$, and Nguyen Ngoc Hai Dang$^\nu$}
    \IEEEauthorblockA{$^{\ast}$School of Information and Communications Technology, Hanoi University of Science and Technology, Vietnam \\
    $^{\xi}$Faculty of Electrical and Electronic Engineering, Hanoi Open University, Vietnam\\
    $^\eta$Posts and Telecommunications Institute of Technology Hanoi, Vietnam\\
    $^\nu$Faculty of Informatics, University of Debrecen, Hungary
    }
}
\maketitle


\begin{abstract}
Mobile edge computing (MEC) allows appliances to offload workloads to neighboring MEC servers that have the potential for computation-intensive tasks with limited computational capabilities. This paper studied how deep reinforcement learning (DRL) algorithms are used in an MEC system to find feasible decentralized dynamic computation offloading strategies, which leads to the construction of an extensible MEC system that operates effectively with finite feedback. Even though the Deep Deterministic Policy Gradient (DDPG) algorithm, subject to their knowledge of the MEC system, can be used to allocate powers of both computation offloading and local execution, to learn a computation offloading policy for each user independently, we realized that this solution still has some inherent weaknesses. Hence, we introduced a new approach for this problem based on the Twin Delayed DDPG algorithm, which enables us to overcome this proneness and investigate cases where mobile users are portable. Numerical results showed that individual users can autonomously learn adequate policies through the proposed approach. Besides, the performance of the suggested solution exceeded the conventional DDPG-based power control strategy.
\end{abstract}


\section{Introduction}

With the rapid expansion of fifth-generation (5G) wireless networks, mobile applications that require intensive computation like geolocation-based augmented and virtual reality (AR/VR), face recognition, or online 3D gaming are challenged by the constrained computational capabilities of mobile devices \cite{6616113}. Mobile edge computing (MEC) \cite{7807196}, whose architecture is shown below in Fig.~\ref{fig:MEC_architecture} has arisen as a hopeful solution to heighten the quality of experience (QoE) for these computationally intensive applications. 

\begin{figure}[htbp]
    \centerline{\includegraphics[width=\linewidth]{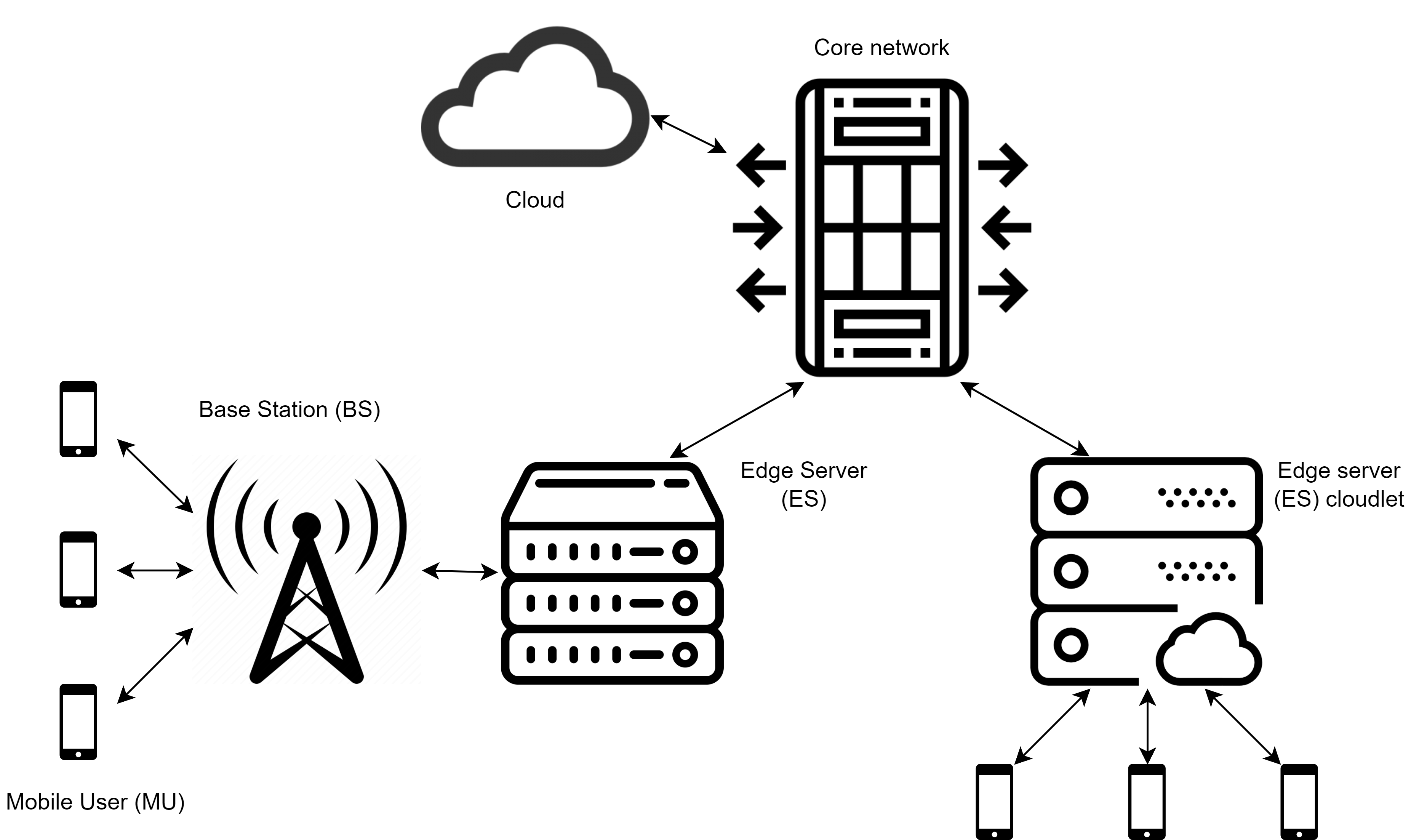}}
    \caption{MEC architecture}
    \label{fig:MEC_architecture}
\end{figure}

MEC aims to focus on the difference between the constrained resources available on mobile devices and the growing computational demands of mobile applications by enabling the offloading of computation workloads to MEC servers connected to a base station (BS). This offloading significantly reduces latency and power consumption, which means that QoE for mobile applications is enhanced. However, task offloading strongly hinges on the wireless transmission's effectiveness, which needs the MEC system to supervise radio resources that come with computational resources and accomplish computation tasks competently. So, researchers have generally and closely examined computation offloading strategies for MEC in the literature to get greater energy potency or computational performance.

Research by \cite{xu2017online} and \cite{8444467} has demonstrated that reinforcement learning (RL) techniques can effectively optimize dynamic computational offloading policies even when the algorithm is free from awareness of the MEC system. Note that traditional RL algorithms have poor scalability in the case of multi-agents because the dimensionality of the state space will increase, thereby making traditional tabular methods infeasible \cite{sutton2018reinforcement}. A new approach, deep reinforcement learning (DRL) is capable of effectively approximating Q-values of RL by making use of deep neural networks (DNNs) for approximation \cite{Mnih2015HumanlevelCT}.

Our study builds upon this foundation by exploiting the DRL framework used in \cite{10.1186/s13638-020-01801-6} where a MIMO-based MEC system that contains a base station, a MEC server and many distinct mobile users with non-static channel condition and random task arrival times is considered. We provide a better method that can give better results for the problem of interest in this framework by using Twin Delayed DDPG (TD3), which is developed from DDPG, used in \cite{10.1186/s13638-020-01801-6}. Another important consideration is we allow mobile users to be portable, which means that the distance between users and the BS may be changed based on their movement. Numerical results for a simulation setup in this case are also shown to prove that the results of TD3 surpass that of DDPG. In Section \ref{sec: MECmodel}, we introduced the MEC network and computation model. In Section \ref{sec: DRL}, we showed the way DRL can be applied in this model. In Section \ref{sec: Experiment}, the superiority of Twin Delayed DDPG was illustrated in a dynamic user setup, and in Section \ref{sec: Conclusion}, we gave the cons.

\section{DYNAMIC COMPUTATION OFFLOADING FOR MEC}
\label{sec: MECmodel}

Our research focuses on a multi-user MEC system. This system comprises three main components: a base station, a MEC server, and a collection of mobile users, denoted as $\mathcal{M} = \{1, 2, \dots, M\}$. Fig. ~\ref{fig:MEC_architecture} illustrates this architecture. Every user $m \in \mathcal{M}$ has computational tasks, but a mobile device’s computation resources are limited, therefore, we need to place the MEC server close to the BS by the telecom operator to promote users’ computational ability by allowing partial offloading of the computational tasks via wireless connections to the MEC server \cite{DBLP:journals/corr/MachB17, Mao2016PowerDelayTI}. The discrete-time model, which is applied later, splits the process into timeslots of equal length $\tau_0$, and we index the time slots by $ \mathcal{T}=\{0,1,\dots\}$. For this system, each user determines the rate of computation offloaded into the BS and computation at local devices at every time slot, as in \cite{9467317}. This decentralized task scheduling is advantageous, particularly if we increase mobile users number because the scheduling reduces system overhead and enhances the MEC system's scalability. Additionally, this approach harmonizes the mean task processing latency and energy consumption, considering that each individual user's channel status and task arrival vary per time slot. The next parts delineate the networking and computing model.
\subsection{Network Model}

A 5G base station that is macro-cell or small-cell with $N$ antennas is considered. By utilizing a zero-forcing (ZF) linear detection algorithm, which has low time complexity and is efficient \cite{9891827}, especially in the context of MIMO for multiple users using large antenna arrays \cite{6457363}, we can handle the uplink transmissions of multiple mobile users where each mobile device has only 1 antenna.

For each $t \in \mathcal{T}$, the received signal of the BS is:
\begin{equation}
    \mathbf{y}(t) = \sum_{j=1}^{M} (\mathbf{h}_m(t) \sqrt{p_{o,m}(t)} s_m(t) + \mathbf{n}(t)) \in \mathbb{C}^{N \times 1}
\end{equation}
where the following definitions hold
\begin{itemize}
    \item $\mathbf{h}_m(t) \in \mathbb{C}^{N \times 1}$: the channel vector of $m \in \mathcal{M}$
    \item $p_{o,m}(t) \in [0, P_{o,m}]$: the transmission power of user $t$ for offloading task data bits. Here, $P_{o,m}$ is the upper limit of $p_{o,m}(t)$.
    \item $s_m(t)$: the complex data symbol with variance $s^2 = 1$
    \item $\mathbf{n}(t) \sim \mathcal{CN}\left(0, \sigma_R^2 \mathbf{I}_N\right)$: an additive white Gaussian noise (AWGN) vector
\end{itemize}

We operate the Gaussian Markov block fading autoregressive model as below to calculate the channel over timeslots for each mobile user $m \in \mathcal{M}$:
\begin{equation}
    \mathbf{h}_m(t) = \rho_m \mathbf{h}_m(t-1) + \sqrt{1 - \rho_m^2(t)} \mathbf{e}(t)
    \label{eq: GMBF_autoreg_model}
\end{equation}
where:
\begin{itemize}
    \item $\rho_m = J_0(2\pi f_{d,m} \tau_0)$: the normalized channel correlation coefficient between timeslots $t-1$ and $t$ in compliance with Doppler spectrum in Jake’s model for Rayleigh fading. Here, $f_{d,m}$ is user $m$'s Doppler frequency, and $J_0$ stands for the Bessel function of the first kind.
    \item $\mathbf{e}(t)$: the complex Gaussian error vector. Note that $\mathbf{e}(t)$ and $\mathbf{h}_m(t)$ are uncorrelated.
\end{itemize}

Let $\mathbf{H}(t) = [\mathbf{h}_1(t), \dots, \mathbf{h}_M(t)]$ be the $N \times M$ channel matrix between $M$ users and the base station. We focus on scenarios where $M < N$, which means the mobile user number is less than the antenna number at the BS. The ZF detector at the base station is: $\mathbf{H}^\dagger(t) = (\mathbf{H}^\mathsf{H}(t)\mathbf{H}(t))^{-1} \mathbf{H}^\mathsf{H}(t)$. In this equation,  $\mathbf{H}^\mathsf{H}$ denotes the Hermitian of $H$. We define $\mathbf{g}_m^\mathsf{H}(t)$ as $\mathbf{H}^\dagger(t)$'s $m$-th row, then for each user $m$, the received signal is:
\begin{equation}
    \mathbf{g}_m^\mathsf{H}(t) \mathbf{y}(t) = \sqrt{p_{0,m}(t)} s_m(t) + \mathbf{g}_m^\mathsf{H}(t) \mathbf{n}(t)
\end{equation}

We can derive this from the fact that in ZF detection, $\mathbf{g}_i^\mathsf{H}(t) \mathbf{h}(t) = \delta_{ij}$. Here, $\delta_{ij}$ is the Kronecker delta.

So, the relevant signal-to-interference-plus-noise (SINR) is computed by:
\begin{equation}
    \gamma_m(t) = \frac{p_{o,m}(t)}{\sigma_R^2 \|\mathbf{g}_m(t)\|^2} = \frac{p_{o,m}(t)}{\sigma_R^2 \left[\left(\mathbf{H}^\mathsf{H}(t)\mathbf{H}(t)\right)^{-1}\right]_{mm}}
    \label{eq: SINR}
\end{equation}

From \eqref{eq: SINR}, it can be confirmed that the SINR of each user deteriorates with an increasing number of users, necessitating users to heighten power allocation to the environment for task offloading based on SINR feedback, which is detailed in the computation model. 

\subsection{Computation Model}

In this subsection, we talk about how a user $m \in \mathcal{M}$ leverages either computation offloading or local execution to meet the requirements of its running applications. Let $a_m(t)$ be the task arrival throughout timeslot $t \in \mathcal{T}$. This can be checked from timeslot $t+1$, with the assumption that the task arrival random variables ${a_i(t)}$ are statistically independent and follow an identical distribution (i.i.d).  In this model, $\lambda_m$ represents the expected value of the random variable for each user $m$, such that $\lambda_m = \mathbb{E}[a_m(t)]$. We presume that computational tasks are divided into two parts: $d_{l,m}(t)$ for local processing and $d_{o,m}(t)$ are transported to the MEC server, indicating that the applications are fine-grained \cite{7264984}. The task buffer queue length for user $m$ at the start of timeslot $t$, denoted as $B_m(t)$, is defined by:
\begin{equation}
    B_m(t+1) = \max\left(B_m(t) - d_{l,m}(t) - d_{o,m}(t), 0\right) + a_m(t)
    \label{eq: task_buffer}
\end{equation}
Here, $B_m(0) = 0$ represents the initial condition.

\subsubsection{Local computing}

For local computation, the scheduled CPU frequency for time slot $t$ can be dynamically adjusted based on the power allocated for local execution, $p_{l,m}(t)$ within the range $[0, P_{l,m}]$. The adjustment process uses Dynamic Voltage and Frequency Scaling (DVFS) techniques \cite{Burd1996ProcessorDF}:
\begin{equation}
    f_m(t) = \sqrt[3]{\frac{p_{l,m}(t)}{\kappa}}
\end{equation}
Here, $\kappa$, which depends on the chip architecture, represents the effective switched capacitance. 

We define the maximum of the CPU-cycle frequency that the device of user $m$ can achieve as:
\begin{equation}
    F_m = \sqrt[3]{\dfrac{P_{l,m}}{\kappa}}
\end{equation}

The computational throughput that user $m$ can process locally during timeslot $t$ is quantified as: 
\begin{equation}
    d_{l,m}(t) = \tau_0 f_m(t) L_m^{-1}
\end{equation}
Here, $L_m$ represents how many CPU cycles are necessary to run a task bit at user $m$. We can use offline measurements to estimate this value\cite{10.5555/1863103.1863107}.

\subsubsection{Edge computing}

In the edge computing scenario, it is crucial to observe that we typically have adequate computational resources for the MEC server, often including high-frequency multi-core CPUs. This allows for the assumption that many tasks can be parallel controlled with the processing latency being eliminated, and negligible feedback delay because of the small size of the computational output. Consequently, we can handle all task data bits transported to the MEC server by way of the base station. So, we can compute how much data user $m$ offloads in timeslot $t$ as:
\begin{equation}
    d_{o,m}(t) = \tau_0 W \log_2(1 + \gamma_m(t))
\end{equation}
Here, $p_{o,m}(t)$ represents the uplink transmission power, $\gamma_m(t)$ is calculated as \eqref{eq: SINR}, and $W$ is the system bandwidth.

\section{DRL BASED DECENTRALIZED DYNAMIC COMPUTATION OFFLOADING}
\label{sec: DRL}
\subsection{DRL Framework}

In \cite{10.1186/s13638-020-01801-6}, in the context of decentralized dynamic computation offloading, Chen and Wang introduced a DRL framework that allows each mobile user to systematically acquire proficient policies for dynamically allocating power between local computation and offloading tasks within a continuous range. Each user lacks prior familiarity with the MEC system, including the statistical information regarding task arrivals and wireless channels, and the total number of users $|\mathcal{M}|$. Hence, the learning process occurs online and is model-free.

\subsubsection{State space}

The system observation contains the task buffer queue lengths and the user channel vectors. However, gathering this information at the BS and allocating it to each user incurs significant system overhead, which rises with the mobile user number. Thus, to enhance the scalability of the MEC system and diminish the expense, each user's state is solely influenced by the system's local observation, and each user independently selects actions.

For a user $m$ at timeslot $t$, the state is represented by:
\begin{equation}
    \mathbf{s}_{m,t} = [B_m(t), \phi_m(t-1), \mathbf{h}_m(t)]
\end{equation}
Here, $\phi_m(t-1)$ represents estimated power ratio after the base station applies zero-forcing detection for the previous timeslot $t-1$, defined by:
\begin{align}
     \phi_m(t) &= \frac{\gamma_m(t) \sigma_R^2}{p_{o,m}(t) \|\mathbf{h}_m(t)\|^2}  \\
     & = \frac{1}{ \left[\left(\mathbf{H}^\mathsf{H}(t)\mathbf{H}(t)\right)^{-1}\right]_{mm}\|\mathbf{h}_m(t)\|^2}
\end{align}

When timeslot $t$ starts, we update the queue length of data buffer $B_m(t)$ for each user $m$. This update follows \eqref{eq: task_buffer}.

\subsubsection{Action Space}

For each user $m$ at timeslot $t$, the action is defined as:
\begin{equation}
    a_{m,t} = [p_{l,m}(t), p_{o,m}(t)]
\end{equation}

This action includes the allocated powers for both computation offloading and local execution, which are selected for each timeslot. Note that the action space can be discrete (using predefined power levels) or continuous, based on the algorithm we use to run, but each element in the action has its bound: $0 \leq p_{l,m}(t) \leq P_{l,m}$, $0 \leq p_{o,m}(t) \leq P_{o,m}$. From that, discrete action spaces' large number of dimensions can be essentially diminished.

\subsubsection{Reward Function}

We aim to develop a dynamic computation offloading strategy for the MEC model, focusing on minimizing energy consumption while ensuring tasks are completed inside an acceptable buffering delay. Therefore, each user's total computation cost is determined by considering both the overall energy consumption and the additional penalty incurred due to task buffering delay. In accordance with Little’s theorem \cite{10.5555/1972549}, the mean task buffer queue length is harmonious with the task buffering delay. Hence, for each user $m$ at timeslot $t$, the reward $r_{m,t}$ is calculated by:
\begin{equation}
r_{m,t} = -w_{m,1} \times (p_{l,m}(t) + p_{o,m}(t)) - w_{m,2}\times B_m(t)
\end{equation}
Here, $w_{m,1}$ and $w_{m,2}$ are defined below, and $r_{m,t}$ is computed by negatively weighting the sum of both the task buffer queue length and total power consumption at timeslot $t$. By adjusting $w_{m,1}$ and $w_{m,2}$, we can express the balance between delay in buffering and energy utilization. The RL algorithms maximizes the value function for user $m$ with initial state $s_{m,1}$ and policy $\mu_m$, and the value function is calculated as:
\begin{equation}
    V^{\mu_m}(\mathbf{s}_{m,1}) = \mathbb{E}\left[\sum_{t=1}^{\infty} \gamma^{t-1} r_{m,t} | \mathbf{s}_{m,1}\right]
\end{equation}

\subsection{DRL algorithm}

\subsubsection{Deep Deterministic Policy Gradient (DDPG)}

This algorithm employs an actor-critic approach with two different DNNs. The first DNN, serving as the actor, approximates the policy network, represented as $\mu(s|\theta^\mu)$. The second DNN, functioning as the critic, estimates the Q-value network, represented as $Q(s, a| \theta^Q)$. Detailed visualization of this algorithm is shown in Fig.~\ref{fig:DDPG_algorithm}:

\begin{figure}[t]
    \centerline{\includegraphics[width=\linewidth]{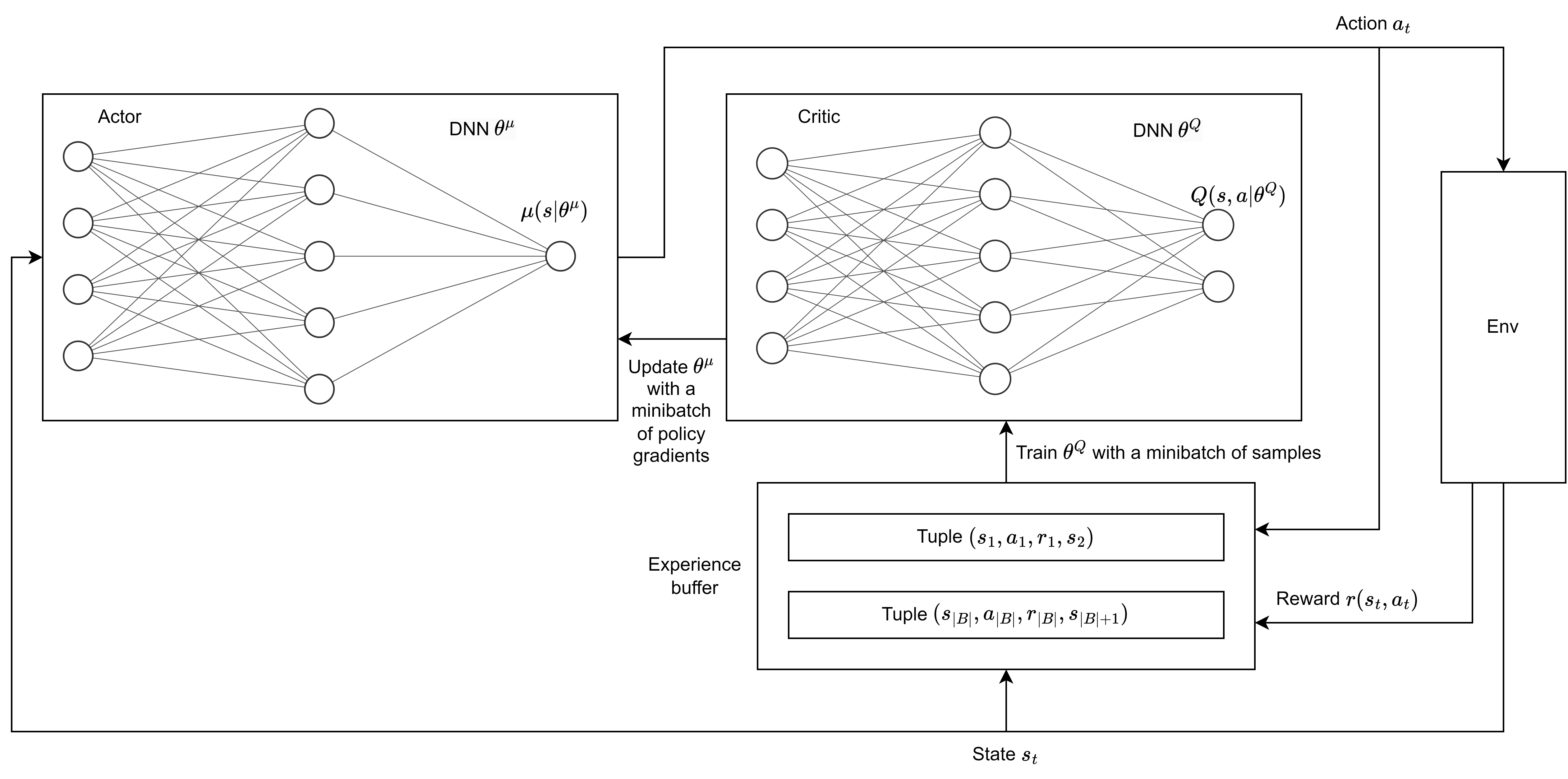}}
    \caption{DDPG Algorithm}
    \label{fig:DDPG_algorithm}
\end{figure}

Using learning rate $\alpha_Q$ for the critic and $\alpha_\mu$ for the actor, in each update step, we update $\theta^Q$ and $\theta^\mu$ by gradient descent (GD) and another minimal parameter called soft update rate $\tau \geq 0$ that accelerates the momentum of $\theta^Q$ and $\theta^\mu$. Note that if we want the algorithm to explore more, we can build the exploration policy. To enhance exploration, the algorithm incorporates a stochastic process. It generates a noise term $\Delta\mu$ and adds it to the actor's output. We clip the result action to get the next action $a'$. Hence, we get the algorithm \ref{algo: DDPG} below:

\begin{algorithm}
\caption{Deep Deterministic Policy Gradient (DDPG)}
\label{algo: DDPG}
\begin{algorithmic}[1]
\State Set up actor parameters $\theta^\mu$, critic parameters $\theta^Q$, and create an empty experience buffer $\mathcal{B}$
\State Equate target parameters to main parameters $\theta^{\mu'} \leftarrow \theta^\mu$, $\theta^{Q'} \leftarrow \theta^Q$
\Repeat
\State Get state $s$ and calculate the action $a = \text{clip}(\mu(s|\theta^\mu) + \epsilon, a_{\text{low}}, a_{\text{high}})$, here, $\epsilon$ is the noise generated by Ornstein-Uhlenbeck process
\State Perform $a$ within the environment
\State Get next state $s'$, reward $r$, and termination signal $d$ indicating if $s'$ is the terminal state
\State Add $(s,a,r,s',d)$ to experience buffer $\mathcal{B}$
\If{$s'$ is the terminal state}
\State Initialize the state of the environment
\EndIf
\If{can update}
\For{num of updates}
\State Select a batch of transitions $B = \{(s,a,r,s',d)\}$ randomly from $\mathcal{B}$
\State Calculate the targets $y(r,s',d) = r + \gamma(1 - d) Q(s',\mu(s'|\theta^{\mu'}) | \theta^{Q'})$
\State Update Q-value network by GD using
\[
\frac{\alpha^Q}{|B|} \nabla_{\theta^Q} \sum_{(s,a,r,s',d) \in B} \left(Q(s,a|\theta^Q) - y(r,s',d)\right)^2
\]
\State Update policy by GD using
\[
\frac{\alpha^\mu}{|B|} \nabla_{\theta^\mu} \sum_{s \in B} Q\left(s,\mu(s|\theta^\mu)|\theta^Q\right)
\]
\State Update target networks with
\[
\theta^{Q'} \leftarrow \tau \theta^Q + (1 - \tau)\theta^{Q'}
\]
\[
\theta^{\mu'} \leftarrow \tau \theta^\mu + (1 - \tau)\theta^{\mu'}
\]
\EndFor
\EndIf
\Until{convergence}
\end{algorithmic}
\end{algorithm}

\subsubsection{Twin Delayed DDPG (TD3)}

Although TD3 also uses DNNs to approximate Q-value and policy networks, by introducing the following important tricks, this algorithm addresses the issue of DDPG when the learned Q-value network starts to overestimate Q-value:

\begin{itemize}
    \item Clipped Double-Q Learning: In the TD3 approach, instead of one Q-value network, two Q-value estimators are learned, and to calculate the target values for the Bellman error loss functions, the smaller one is used:
    \begin{equation}
        y(r, s', d) = r + \gamma (1 - d) \min_{i=1,2} Q(s, a'(s') | \theta^{Q_i})
    \end{equation}
    The training process for both Q-value networks uses this target, thereby mitigating overestimation in the network.
    \item “Delayed” Policy Updates: TD3 implements a delayed update mechanism for its policy network. This approach involves less frequent updates to the policy compared to the Q-value networks, which helps mitigate the instability that typically occurs in DDPG.
\end{itemize}

Given these modifications, the pseudocode of the TD3 algorithm is shown in Algorithm 1. Although we simultaneously update 2 Q-value networks, we lower the number of updates of the policy and target networks, which ensures that the runtime of Twin Delayed DDPG isn't much more than that one of DDPG. In the following section, the simulation setup and experimental results of two algorithms DDPG and TD3 in a multi-user environment, which is almost the same as in \cite{10.1186/s13638-020-01801-6}, are discussed.

\begin{algorithm}
\caption{Twin Delayed DDPG}
\label{algo: TD3}
\begin{algorithmic}[1]
\State Set up actor parameters $\theta^\mu$, critic parameters $\theta^Q$, error range parameter $c$, policy delay parameter \texttt{update\_every} and create an empty experience buffer $\mathcal{B}$.
\State Equate target parameters to main parameters $\theta^{\mu'} \leftarrow \theta^\mu$, $\theta^{Q'_1} \leftarrow \theta^{Q_1}$, $\theta^{Q'_2} \leftarrow \theta^{Q_2}$
\Repeat
\State Get state $s$ and calculate the action $a = \text{clip}(\mu(s|\theta^\mu) + \epsilon, a_{\text{low}}, a_{\text{high}})$, here, $\epsilon$ is noise generated by Ornstein-Uhlenbeck process
\State Perform $a$ within the environment
\State Get next state $s'$, reward $r$, and termination signal $d$ indicating if $s'$ is the terminal state
\State Add $(s,a,r,s',d)$ to experience buffer $\mathcal{B}$
\If{$s'$ is the terminal state}
\State Initialize the state of the environment
\EndIf
\If{can update}
\For{$j$ in num of updates}
\State Select a batch of transitions $B = \{(s,a,r,s',d)\}$ randomly from $\mathcal{B}$
\State Calculate the next action $a' = \text{clip}(\mu(s'|\theta^\mu) + \text{clip}(\epsilon, -c, c), a_{\text{low}}, a_{\text{high}})$ from $\mathcal{B}$
\State Calculate the targets $y(r,s',d) = r + \gamma(1-d)\min_{i=1,2} Q(s',a'|\theta^{Q'_i})$
\State Update Q-value network by GD for $i = 1, 2$ using
\[
\frac{\alpha^Q}{|B|} \nabla_{\theta^{Q_i}} \sum_{(s,a,r,s',d) \in B} \left(Q(s,a|\theta^{Q_i}) - y(r,s',d)\right)^2
\]
\If{$j\mod$\texttt{update\_every} $= 0$}
\State Update policy by GD using
\[
\frac{\alpha^\mu}{|B|} \nabla_{\theta^\mu} \sum_{s \in B} Q\left(s,\mu(s|\theta^\mu)|\theta^{Q_1}\right)
\]
\State Update target networks with
\[
\theta^{Q'_i} \leftarrow \tau \theta^{Q_i} + (1 - \tau)\theta^{Q'_i} \text{ for } i = 1,2
\]
\[
\theta^{\mu'} \leftarrow \tau \theta^\mu + (1 - \tau)\theta^{\mu'}
\]
\EndIf
\EndFor
\EndIf
\Until{convergence}
\end{algorithmic}
\end{algorithm}

\section{Experimental results}
\label{sec: Experiment}
\subsection{Simulation setup}

In our MEC system, we set the timeslot duration to $\tau_0 = 1$ms long. We initially set $\mathbf{h}_m(0) \sim \mathcal{CN}\left(0, h_0 \left(\dfrac{d_0}{d_m}\right)^\alpha \mathbf{I}_N\right)$ as the channel vector for each user $m$ where:
\begin{itemize}
    \item $h_0 = -30$dB: the path-loss constant
    \item $d_0 = 1$m: the reference distance
    \item $d_m$: the actual distance between user $m$ (m) and the base station
    \item $\alpha = 3$: the path-loss exponent
\end{itemize}

\begin{figure}[t]
    \centerline{\includegraphics[width=\linewidth]{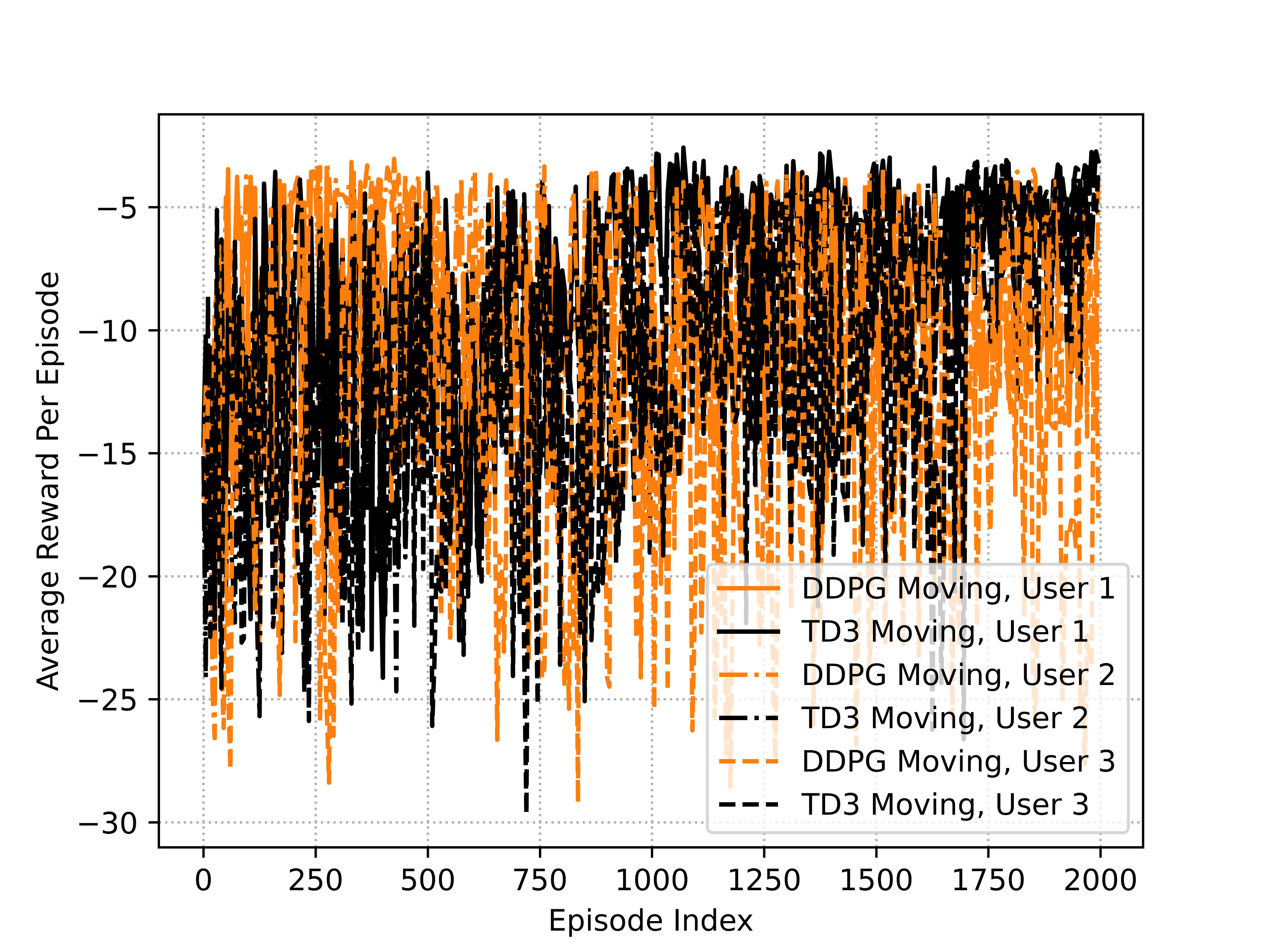}}
    \caption{Average reward per episode}
    \label{fig:avg_reward}
\end{figure}

\begin{figure}[htbp]
    \centerline{\includegraphics[width=\linewidth]{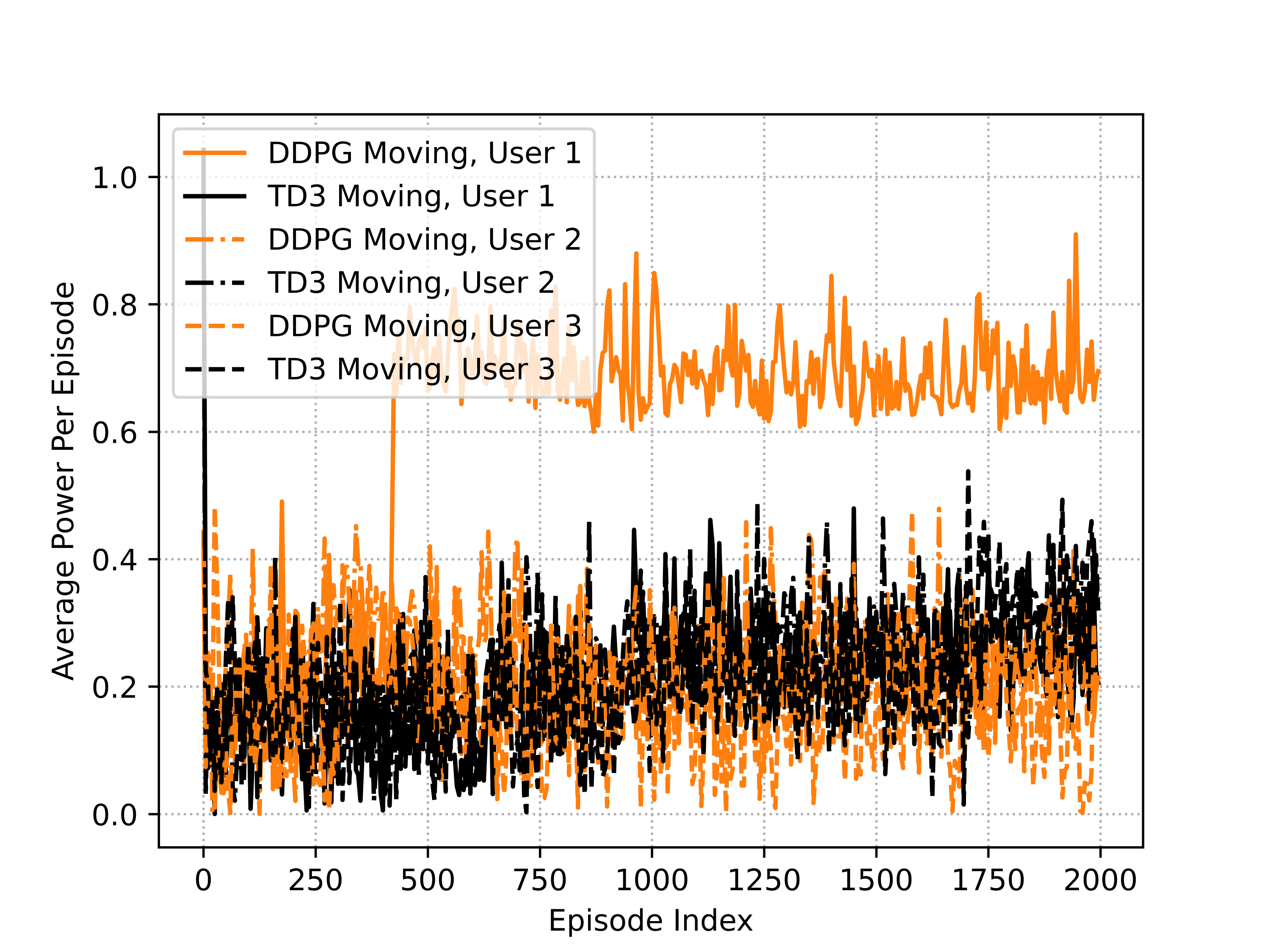}}
    \caption{Average power per episode}
    \label{fig:avg_power}
\end{figure}

\begin{figure}[t]
    \centerline{\includegraphics[width=\linewidth]{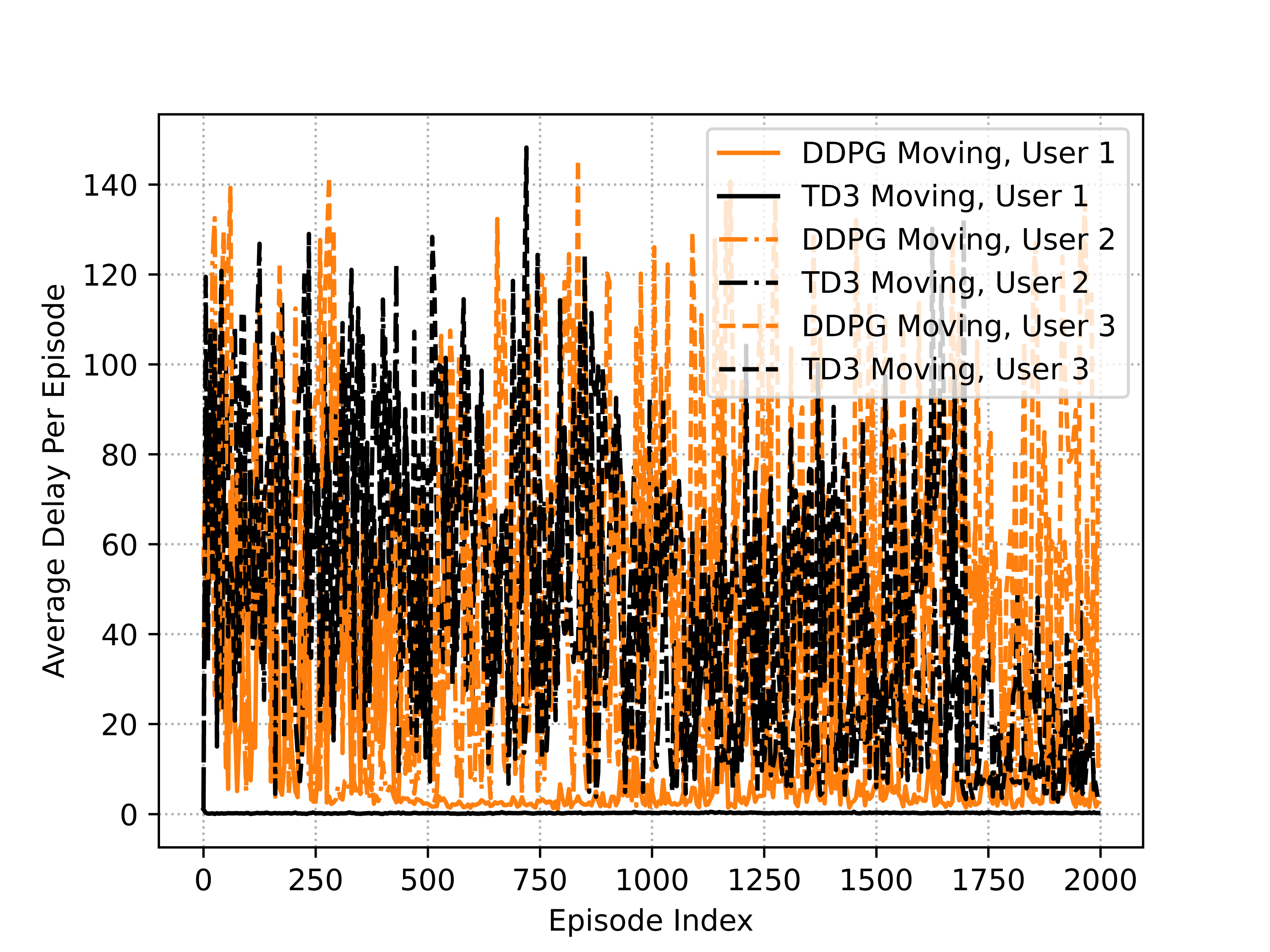}}
    \caption{Average delay per episode}
    \label{fig:avg_delay}
\end{figure}

After each slot, the channel $\mathbf{h}_m(t)$ updates by \eqref{eq: GMBF_autoreg_model}. We set the correlation coefficient $\rho_m$ to $0.95$ and the Doppler frequency of user $m$ $f_{d,m}$ to $70$Hz. We initially set $\mathbf{e}(t) \sim \mathcal{CN}\left(0, h_0 \left(\dfrac{d_0}{d}\right)^\alpha \mathbf{I}_N\right)$ as the error vector. Moreover, we use a system bandwidth $W$ of $1$MHz, and we set the noise power $\sigma_R^2 = 10^{-9}$W and the maximum transmission power $P_{o,m} = 2$W. Conversely, for local processing, we set the maximum local execution power for each user $P_{l,m}$ to $2$W, and the effective switched capacitance $\kappa$ to $10^{-27}$. We can derive the maximum CPU-cycle frequency for each user's device from them and get $F_m = 1.26$GHz. One more value to consider is that we need $L_m = 500$ CPU cycles per task bit for every user.

In this system, there are $M = 3$ mobile users, each randomly positioned $d_m = 100$m away from the base station. We assign different task arrival rates to each user. Specifically, it is $\lambda_m = 1.2 - 0.1m$ Mbps for user $m \in \{1,2,3\}$, which is different from the task arrival rate in \cite{10.1186/s13638-020-01801-6}.

To implement the DRL algorithms, every user agent $m$ utilizes a densely connected neural network for both the actor and critic network. For the neural networks, we use almost the same settings as in \cite{10.1186/s13638-020-01801-6}. Network layer weights are initialized following the settings outlined in the experiment of \cite{Lillicrap2015ContinuousCW}. To facilitate action exploration, we apply the Ornstein-Uhlenbeck process ($\theta = 0.15$, $\sigma = 0.12$), and we maintain an experience replay buffer of $|\mathcal{B}| = 2.5\times 10^5$.

For each user agent $m$, the reward function $r_{m,t}$ is expressed as:
\begin{equation}
    r_{m,t} = -w_{m,1} (p_{l,m}(t) + p_{o,m}(t)) - w_{m,2} B_m(t)
\end{equation}

Here, two non-negative weighting factors $w_{m,1}, w_{m,2}$ are computed using a balance factor $w_m \in [0,1]$ for every user agent $m \in \mathcal{M}$: $w_{m,1} = 10w_m, w_{m,2} = 1 - w_m$. Furthermore, we run $K_{\max} = 2000$ episodes, and each episode has at most $T_{\max} = 200$ steps in the experiment. We set $w_m = 0.8$ for all users. To represent users' movement, for each state transition in each episode, the users' distances to the BS are added by a value drawn randomly from the standard normal distribution, such that the absolute value of the cumulative sum of these distances is lower than 10.

\subsection{Experimental results}

The results when all episodes have ended are displayed in Fig.~\ref{fig:avg_reward}, \ref{fig:avg_power} and \ref{fig:avg_delay}, where orange lines are of DDPG, and black lines are of TD3, and another result table for all mobile users are demonstrated in Table \ref{tab: results}.

\begin{table}[t]
\caption{RESULTS}
\begin{center}
\resizebox{\columnwidth}{!}{%
\begin{tabular}{|c|c|c|c|c|c|c|}
\hline
& \multicolumn{2}{c|}{\textbf{Average reward ($\uparrow$)}} & \multicolumn{2}{c|}{\textbf{Average power ($\downarrow$)}} & \multicolumn{2}{c|}{\textbf{Average delay ($\downarrow$)}} \\
\hline
& DDPG & TD3 & DDPG & TD3 & DDPG & TD3\\
\hline
\textbf{User 1} & -9.52 & \textbf{-3.09} & 0.70 & \textbf{0.21} & \textbf{1.95} & 4.67 \\
\hline
\textbf{User 2} & -14.08 & \textbf{-3.71} & 0.27 & \textbf{0.24} & 59.47 & \textbf{8.91} \\
\hline
\textbf{User 3} & -19.49 & \textbf{-3.89} & \textbf{0.20} & 0.30 & 89.30 & \textbf{5.29} \\
\hline
\end{tabular}%
}
\label{tab: results}
\end{center}
\end{table}

In \cite{10.1186/s13638-020-01801-6}, DDPG demonstrates superiority over other algorithms in terms of average reward per episode and average power per episode, and our experiment has revealed that TD3 has much better results than DDPG in these aspects. In addition, regarding average power per episode, TD3 also achieved better results compared with DDPG.

\section{Conclusion}
\label{sec: Conclusion}
The paper introduced a new method that leverages deep reinforcement learning to find a dynamic computation offloading strategy in mobile edge computing. The proposed Twin Delayed DDPG algorithm outperformed the previous DDPG algorithm, particularly when users are portable. In the future, we will consider many other simulation setups and use Multi-agent Reinforcement Learning algorithms to provide better results for this problem.

\bibliographystyle{unsrt}
\bibliography{bibliography}

\end{document}